\documentclass[reprint,twocolumn]{revtex4}%
\usepackage{amsfonts}
\usepackage{amsmath}
\usepackage{amssymb}
\usepackage{charter}
\usepackage{graphicx}%
\providecommand{\U}[1]{\protect \rule{.1in}{.1in}}
\begin{document}
\title{Optimal fidelity of teleportation with continuous-variable using three-tunable
parameters in realistic environment}
\author{{\small Li-Yun Hu}$^{1,2\ast}$\thanks{Corresponding author. Email:
hlyun2008@126.com}, Zeyang Liao$^{1}$, Shengli Ma$^{1}$ and {\small M. Suhail
Zubairy}$^{1\ast}$}
\affiliation{$^{1}${\small Institute for Quantum Science and Engineering (IQSE) and
Department of Physics and Astronomy, Texas A\&M University, College Station,
TX 77843, USA}}
\affiliation{$^{2}${\small Center for Quantum Science and Technology, Jiangxi Normal
University, Nanchang 330022, China}}
\affiliation{*Corresponding author.}

\begin{abstract}
We introduce three tunable parameters to optimize the fidelity of quantum
teleportation with continuous-variable in nonideal scheme. Using the
characteristic function formalism, we present the condition that the
teleportation fidelity is independent of the amplitude of input coherent
states for any entangled resource. Then we investigate the effects of tunable
parameters on the fidelity with or without the presence of environment and
imperfect measurements, by analytically deriving the expression of fidelity
for three different input coherent state distributions. It is shown that, for
the linear distribution, the optimization with three tunable parameters is the
best one with respect to single- and two-parameter optimization. Our results
reveal the usefulness of tunable parameters for improving the fidelity of
teleportation and the ability against the decoherence.

{\small PACS number(s): 03.65 -a. 42.50.Dv}

\end{abstract}
\maketitle

\section{Introduction}

Quantum teleportation has an indispensable role in the manipulation of quantum
states and processing of quantum information \cite{1,2,3,3a}. Usually, the
two-mode squeezed vacuum is often used as the entanglement resource with
continuous-variable (CV). However, due to the limit of experiment, it is hard
to achieve high degree of squeezing which leads to a low effect of
teleportation fidelity.

In order to increase the entanglement and fidelity of teleportation, a number
of stratigies have been proposed \cite{4,5,6,7,8,9,10,11,12,13}. Among them,
non-Gaussian operations including the photon subtraction $a$ or addition
$a^{\dag}$ or the superposition of both can be used to realize this purpose
for given Vaidman-Brauntein-Kimble (VBK) scheme. For example, the
superposition\ operator $ta^{\dag}+ra$ is proposed for quantum state
engineering to transform a classical state to a nonclassical one \cite{14},
and then is performed on two-mode squeezed vacuum (TMSV) for enhancing quantum
entanglement as well as the fidelity of teleportation \cite{15}. It is found
that the fidelity teleporting coherent state can be further improved by
optimizing the superposition operation compared with the other non-Gaussian
states, such as photon-subtraction TMSV. As another example, a remarkable
improvement of the teleportation fidelity with CV can be obtained by
introducing another optimal non-Gaussian resources when given the usual and
(non-)ideal VBK scheme \cite{16,17,18} by considering imperfect Bell
measurements and damping. In Ref. \ref{17}, the \textquotedblleft
shot-fidelity\textquotedblright \ and single-gain factor are used to discuss
the performance of teleportation. In fact, these protocols above are employed
to enhance the fidelity of teleportation by changing quantum entangled resources.

Then an inverse question is that given a certain class of entangled resources
with some given properties, how we can modify the VBK scheme to improve the
fidelity of teleportation. It is interesting to notice that there is an
alternative method to improve the fidelity of teleportation by using calssical
information. For instance, the teleportation fidelity of CV can be enhanced by
tuning the gain in the measurement dependent modulation on the output field
\cite{6,19} in Heisenberg picture and EPR resources without loss, which has
been realized experimentally by Furusawa et al \cite{20}. However, these two
important theoretical works are concerned with the study of the ideal protocol
implementation using Gaussian resources \cite{6,19}. In addition, there are
some other strategies of gain tuning and of optimal gain \cite{21,22}, using
the Heisenberg picture and the Wigner function, but they can not be directly
applied to more general cases. In addition, in Refs. \cite{21,23} the gain
factor is used to maximize the teleportation fidelity for the case of Gaussian
resources, but the gain-optimized fidelity of teleportation is strongly
suppressed when considered the dissipation. Recently, a hybrid entanglement
swapping protocol is proposed experimentally to transfer discrete-variable
(DV) entanglement by using continuous-variable (CV) Gaussian entangled
resources and by tuning a gain factor of the teleporter \cite{22,24,25}, which
shows that DV entanglement remains present after teleportation for any
squeezing by optimal gain tuning. For more information about advances in
quantum teleportation, we refer to a recent review paper \cite{26} and
references therein.

The fidelity of teleportation, as mentioned above, can be improved by using
tunable entangled resources or classical parameters \cite{16,17,19,22}. In
Ref. \cite{19}, a three-parameter optimal stratigy is introduced to improve
the quality of teleporttaion, including unbalanced beam splitter (BS) and two
non-unity gains. However, they only considered an ideal case. Actually, the
interaction between quantum system and environment can not be avoided and Bell
measurements are usually imperfect. Thus, it would be interesting that whether
it is still effective to enhance the fidelity by using these tunable
parameters in realistic case. In this paper, using the characteristics
function (CF) formalism, we shall extend the analysis of the parameter
optimization strategy for realistic input states and non-ideal entangled
resources and investigate the nonideal quantum teleportation by deriving an
analytical expression of the teleportation fidelity. This formalism is very
convenient to discuss the teleportation for the nonideal case and any
entangled resources. As far as we know, there is no related report up to now.

This paper is arranged as following. In Sec. II, we give a description of the
characteristic function formulism for the case of nonideal parameterized
teleportation with CV scheme. In this scheme, we find the condition that the
fidelity is independent of the amplitude of input coherent states for any
entangled resource. And then we present a qualitative description about
fidelity and average fidelity. In Sec. III, we derive the analytical
expression of the fidelity of teleportation when the TMSV and coherent states
are used as entangled channel and teleported states, respectively. In Sec. IV
we study the performance of amplitude-independent optimal fidelity using the
condition found in Sec. II. It is found that the optimal condition is just
that the two-gain factors and $\theta$ are equal to unit and $\pi/4$,
respectively. Sec. V is devoted to discussing the optimal fidelity over these
three tunable parameters and three different probability distributions for the
input coherent states by deriving the analytical expression of the optimal
fidelity. Our conclusions are drawn in the last section.

\section{Mode and quantitative analysis}

Here, we consider a more realistic case of teleportation scheme shown in Fig.
1(a). In this scheme, there are three tunable parameters, unbanlanced BS and
two non-unity gains ($g_{q}$ and $g_{p}$). The input state (mode 1) and the
entangled resources (shared by modes 2 and 3) are not limited to be pure
states. Considering that the mode 2 can be prepared close to the sender Alice
while the mode 3 usually has to propagate over much longer distance, we can
assume that the mode 2 is not affected by losses but the mode 3 is. In
addition, two symmetrical lossy bosonic channel have been considered before
making Bell measurements, which are simulated through an extra vacuum mode and
a beam splitter with transmission coefficient $T$. The input states of modes 4
and 5 are pure vacuum states.

\begin{figure}[ptb]
\label{Fig0}
\centering \includegraphics[width=0.8\columnwidth]{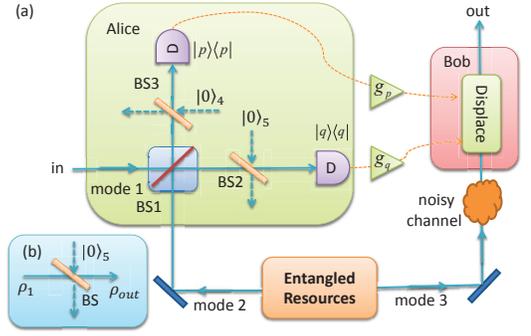}\caption{(Colour
online) Realistic schematic diagram for CV teleportation. BS: Beam splitter.}%
\end{figure}

Next, we shall give a description about the scheme in the formalism of CF
where it is very convenient to discuss the teleportation for the nonideal case
and non-Gaussian entangled resourecs \cite{17,27}.

\subsection{The input-output relation of BS in CF formalism}

In order to obtain the relation between input and output, we first calculate
the output of a beam splitter with a vacuum and an arbitrary density operator
as inputs shown in Fig. 1(b). For simplicity, we denote the vacuum and the
input state as $\left \vert 0\right \rangle _{5}$\ and $\rho_{1}$, respectively.
The output state (denoted as $\rho_{1}^{\prime}$) is given by
\begin{equation}
\rho_{1}^{\prime}=\text{Tr}_{5}\left[  B_{15}\left(  T\right)  \rho_{1}%
\otimes \left \vert 0\right \rangle _{5,5}\left \langle 0\right \vert B_{15}^{\dag
}\left(  T\right)  \right]  ,\label{1}%
\end{equation}
where Tr$_{5}$ is the partial trace over the ancilla mode 5 and $B_{kl}\left(
T\right)  =\exp[\varphi(a_{k}a_{l}^{\dag}-a_{k}^{\dag}a_{l})]$ is the beam
splitter operator describing the interaction between modes 1 and 5 with
$\cos \varphi=\sqrt{T}$ and $a_{k,l}$ ($k=1,l=5$) being the photon annihilation
operator of the $k(l)-$modes. Using the Weyl expansion of density operator, we
can express the density operator $\rho_{1}$ and the vacuum projector
$\left \vert 0\right \rangle _{55}\left \langle 0\right \vert $ as the following
forms:
\begin{align}
\rho_{1} &  =%
{\displaystyle \int}
\frac{d^{2}\alpha}{\pi}\chi_{1}\left(  \alpha \right)  D_{1}\left(
-\alpha \right)  ,\nonumber \\
\left \vert 0\right \rangle _{55}\left \langle 0\right \vert  &  =%
{\displaystyle \int}
\frac{d^{2}\beta}{\pi}e^{-\frac{1}{2}\left \vert \beta \right \vert ^{2}}%
D_{5}\left(  -\beta \right)  ,\label{2}%
\end{align}
where $D_{1}\left(  \alpha \right)  =\exp \{ \alpha a_{1}^{\dag}-\alpha^{\ast
}a_{1}\}$ is the displacement operator, and $\chi_{1}\left(  \alpha \right)  $
is the CF of $\rho_{1}$. On the other hand, using the following transformation
relation
\begin{equation}
B_{15}D_{1}\left(  -\alpha \right)  D_{5}\left(  -\beta \right)  B_{15}^{\dag
}=D_{1}\left(  \bar{\alpha}\right)  D_{5}\left(  \bar{\beta}\right)
,\label{3}%
\end{equation}
where $R=1-T$, and $\bar{\alpha}=\beta \sqrt{R}-\alpha \sqrt{T},\bar{\beta
}=-\beta \sqrt{T}-\alpha \sqrt{R}$, we can derive
\begin{align}
&  \text{Tr}_{5}\left[  B_{15}D_{1}\left(  -\alpha \right)  D_{5}\left(
-\beta \right)  B_{15}^{\dag}\right]  \nonumber \\
&  =D_{1}\left(  \bar{\alpha}\right)  \text{Tr}_{5}\left[  D_{5}\left(
\bar{\beta}\right)  \right]  \nonumber \\
&  =D_{1}\left(  \bar{\alpha}\right)  \pi \delta^{(2)}(\bar{\beta}).\label{3a}%
\end{align}
Here we have used the relation Tr$_{5}D_{5}(\bar{\beta})=\pi \delta^{(2)}%
(\bar{\beta})$. Then substituting Eqs. (\ref{2}) and (\ref{3a}) into Eq.
(\ref{1}) yields
\begin{align}
\rho_{1}^{\prime} &  =%
{\displaystyle \int}
\frac{d^{2}\alpha d^{2}\beta}{\pi^{2}}e^{-\frac{1}{2}\left \vert \beta
\right \vert ^{2}}\chi_{1}\left(  \alpha \right)  \text{Tr}_{5}\left[
D_{1}\left(  \bar{\alpha}\right)  D_{5}\left(  \bar{\beta}\right)  \right]
\nonumber \\
&  =%
{\displaystyle \int}
\frac{d^{2}\alpha d^{2}\beta}{\pi^{2}}e^{-\frac{1}{2}\left \vert \beta
\right \vert ^{2}}\chi_{1}\left(  \alpha \right)  D_{1}\left(  \bar{\alpha
}\right)  \pi \delta^{(2)}\left(  \bar{\beta}\right)  \nonumber \\
&  =%
{\displaystyle \int}
\frac{d^{2}\alpha}{\pi}e^{-\frac{1}{2}R\left \vert \alpha \right \vert ^{2}}%
\chi_{1}\left(  \sqrt{T}\alpha \right)  D_{1}\left(  -\alpha \right)  ,\label{4}%
\end{align}
where $e^{-\frac{1}{2}\left \vert \beta \right \vert ^{2}}$ is the CF of the
vacuum state, and in the second step in Eq. (\ref{4}) the CF $\chi_{1}\left(
\alpha \right)  $ is transformed to $\chi_{1}(\sqrt{T}\alpha)$ with a Gaussian
term $e^{-\frac{1}{2}R\left \vert \alpha \right \vert ^{2}}$ due to the photon
loss. It is then convenient to obtain the input-output relation of the
teleportation scheme shown in Fig. 1(a) as follows.

\subsection{The input-output relation of the teleportation scheme in CF
formalism including photon-loss or imperfect Bell measurements}

Now, we consider the effect of photon-loss on the relation between input and
output of the teleportation scheme in CF formalism. Here we use BS2 and BS3
with vacuum inputs to simulate the photon loss or imperfect Bell measurements
(see Fig. 1(b)), and denote the teleported state, entangled resource, and
auxiliary vacuum as $\rho_{1},$ $\rho_{23}$ and $\left \vert 00\right \rangle
_{45}$, respectively. In oder to realize the teleportation, Alice shall make
her Bell measurements. Before she does, the unitary state evolution can be
formulated as%
\begin{equation}
\rho_{1-5}=U\otimes \rho_{1}\otimes \rho_{23}\otimes \left \vert 00\right \rangle
_{45}\left \langle 00\right \vert U^{\dagger},\label{5}%
\end{equation}
where the unitary evolution operator is defined as $U\left.  =\right.
B_{24}B_{15}B_{12}$ and $B_{kl}$ are the BS operator defined before. In a
similar way to deriving Eq. (\ref{4}), and using the Weyl expansion for the
entangled resource, the reduced output state denoted as $\rho_{1-3}\equiv
$Tr$_{45}\rho_{1-5}$ is given by
\begin{align}
\rho_{1-3} &  =\text{Tr}_{45}\left[  U\otimes \rho_{1}\otimes \rho_{23}%
\otimes \left \vert 00\right \rangle _{45}\left \langle 00\right \vert U^{\dagger
}\right]  \nonumber \\
&  =%
{\displaystyle \int}
\frac{d^{2}\alpha d^{2}\beta d^{2}\gamma}{\pi^{3}}\chi_{1}\left(
\alpha \right)  \chi_{23}\left(  \beta,\gamma \right)  \nonumber \\
&  \times \text{Tr}_{45}\left[  UD_{1}\left(  -\alpha \right)  D_{2}\left(
-\beta \right)  D_{3}\left(  -\gamma \right)  \left \vert 00\right \rangle
_{45}\left \langle 00\right \vert U^{\dagger}\right]  \nonumber \\
&  =%
{\displaystyle \int}
\frac{d^{2}\alpha d^{2}\beta d^{2}\gamma}{\pi^{3}}\chi_{1}\left(
\alpha \right)  \chi_{23}\left(  \beta,\gamma \right)  \text{Tr}_{45}%
[B_{24}\nonumber \\
&  \times B_{15}D_{1}(-\alpha_{1})D_{2}(-\beta_{1})D_{3}\left(  -\gamma
\right)  \left \vert 00\right \rangle _{45}\left \langle 00\right \vert
B_{24}^{\dagger}B_{15}^{\dagger}],\label{6}%
\end{align}
where $B_{12}D_{1}\left(  -\alpha \right)  D_{2}\left(  -\beta \right)
B_{12}^{\dag}=D_{1}(-\alpha_{1})D_{2}(-\beta_{1})$ with $\alpha_{1}=\alpha
\cos \theta-\beta \sin \theta$, $\beta_{1}=\beta \cos \theta+\alpha \sin \theta$ and
$\cos^{2}\theta$ being the transmission coefficient of beam splitter $B_{12}$.
Using Eq. (\ref{1})\ and (\ref{3a}), we can obtain%
\begin{align}
\rho_{1-3} &  =%
{\displaystyle \int}
\frac{d^{2}\alpha d^{2}\beta d^{2}\gamma}{\pi^{3}}\chi_{1}(\sqrt{T}\alpha
)\chi_{23}(\sqrt{T}\beta,\gamma)\nonumber \\
&  \times e^{-\frac{R}{2}(\left \vert \alpha_{1}\right \vert ^{2}+\left \vert
\beta_{1}\right \vert ^{2})}D_{1}\left(  -\alpha_{1}\right)  D_{2}\left(
-\beta_{1}\right)  D_{3}\left(  -\gamma \right)  .\label{7}%
\end{align}
Eq. (\ref{7}) is the representation in CF of deduced density operator before
Bell measurements but after BS2 and BS3.

Then, as the first step of teleportation, Alice makes a joint measurements for
modes 1 and 2 at the output ports, i.e., measures two observables
corresponding to coordinate and momentum of modes 1 and 2. Thus after the
measurements, the outcomes $\rho_{M}$ ($M$ means measurement) in mode 3 are%
\begin{equation}
\rho_{M}\equiv \frac{1}{P\left(  q,p\right)  }\text{Tr}_{12}\left[  \left \vert
q\right \rangle _{11}\left \langle q\right \vert \otimes \left \vert p\right \rangle
_{22}\left \langle p\right \vert \rho_{1-3}\right]  , \label{8}%
\end{equation}
where $P\left(  q,p\right)  $ is the probability distribution function of the
Bell measurement outcomes, $P\left(  q,p\right)  =$Tr$_{3}\{$Tr$_{12}%
[\left \vert q\right \rangle _{11}\left \langle q\right \vert \otimes \left \vert
p\right \rangle _{22}\left \langle p\right \vert \rho_{1-3}]\}$, and $\left \vert
q\right \rangle _{1}\ $and $\left \vert p\right \rangle _{2}$ are the eigenstates
of coordinate and momentum operators $Q_{1}$ and $P_{2}$ corresponding to
modes 1 and 3, respectively.

According to the definition of CF, and using the following relations%
\begin{align}
\text{Tr}_{1}\left[  \left \vert q\right \rangle _{1,1}\left \langle q\right \vert
D_{1}\left(  \alpha \right)  \right]   &  =e^{i\sqrt{2}q\operatorname{Im}%
\alpha}\delta \left(  \sqrt{2}\operatorname{Re}\alpha \right)  ,\nonumber \\
\text{Tr}_{2}\left[  \left \vert p\right \rangle _{2,2}\left \langle p\right \vert
D_{2}\left(  \beta \right)  \right]   &  =e^{-i\sqrt{2}p\operatorname{Re}\beta
}\delta \left(  \sqrt{2}\operatorname{Im}\beta \right)  , \label{9}%
\end{align}
and
\begin{equation}
\text{Tr}_{3}\left[  D_{3}\left(  -\gamma \right)  D_{3}\left(  \eta \right)
\right]  =\pi \delta^{(2)}\left(  \eta-\gamma \right)  , \label{10}%
\end{equation}
the CF of $\rho_{M}$ defined as $\chi_{M}\left(  q,p;\eta \right)  $%
=Tr$_{3}\left[  \rho_{M}D_{3}\left(  \eta \right)  \right]  $ reads%

\begin{align}
&  \chi_{M}\left(  q,p;\eta \right) \nonumber \\
&  =\frac{P^{-1}\left(  q,p\right)  }{\sin2\theta}%
{\displaystyle \int}
\frac{d^{2}\alpha}{\pi^{2}}\exp \left \{  \alpha^{\ast}\xi-\alpha \xi^{\ast
}\right \} \nonumber \\
&  \times \chi_{1}(\sqrt{T}\alpha)\chi_{23}\left[  \sqrt{T}\left(  \alpha
\cot2\theta+\alpha^{\ast}\csc2\theta \right)  ,\eta \right] \nonumber \\
&  \times \exp \left \{  -\frac{R}{2}[(\operatorname{Re}\alpha)^{2}\csc^{2}%
\theta+(\operatorname{Im}\alpha)^{2}\sec^{2}\theta]\right \}  , \label{11}%
\end{align}
where we have defined $\xi=(q/\cos \theta+ip/\sin \theta)/\sqrt{2}$. When $T=1$,
Eq. (\ref{11}) just reduces to Eq.(2.9) in Ref. \cite{27}.

After Alice informs the measured results ($q,p$) to Bob, Bob needs to make a
unitary transformation to obtain the output state at this stage. Here, we
consider the unitary transformation to be the displacement operator
$D_{3}\left(  Z\right)  $ with non-nunity and aysmetrical gains
characteristic, where $Z\equiv g_{q}q+ig_{p}p$ with $g_{q}$ and $g_{p}$ are
two tunable gain parameters. Thus, after the displacement, the output state
can be expressed as $\rho_{D}\equiv \int d^{2}\eta \chi_{M}\left(
q,p;\eta \right)  D_{3}\left(  Z\right)  D_{3}\left(  -\eta \right)
D_{3}\left(  -Z\right)  /\pi$. Usually, we do not have interest in every
measurement result but the average effect. Thus we perform an ensemble
averaging over all measurement results, then the average CF $\bar{\chi}_{out}$
is given by%
\begin{align}
\bar{\chi}_{out}\left(  \beta \right)   &  =\text{Tr}_{3}[D_{3}\left(
\beta \right)  \int dqdpP\left(  q,p\right)  \rho_{D}]\nonumber \\
&  =\chi_{1}\left(  f_{1}\beta-f_{2}\beta^{\ast}\right)  \chi_{23}\left(
\beta^{\ast}f_{3}-\beta \allowbreak f_{4},\beta \right) \nonumber \\
&  \times \exp \{-R[g_{p}^{2}(\operatorname{Re}\beta)^{2}+g_{q}^{2}%
(\operatorname{Im}\beta)^{2}]\}, \label{12}%
\end{align}
where
\begin{align}
f_{1}  &  =\sqrt{\frac{T}{2}}\left(  g_{q}\cos \theta+g_{p}\sin \theta \right)
,\nonumber \\
f_{2}  &  =\sqrt{\frac{T}{2}}\left(  g_{q}\cos \theta-g_{p}\sin \theta \right)
,\nonumber \\
f_{3}  &  =\sqrt{\frac{T}{2}}\left(  g_{q}\sin \theta+g_{p}\cos \theta \right)
,\nonumber \\
f_{4}  &  =\sqrt{\frac{T}{2}}\left(  g_{q}\sin \theta-g_{p}\cos \theta \right)  ,
\label{13}%
\end{align}
and $T$ and $R$ ($T+R=1$) denote, respectively, the transmissivity and
reflectivity of the BS2 and BS3 that stimulate the photon losses.

\subsection{Relation between input and output including noise in mode 3}

Next, we consider the effect in CF formalism of decoherence of environment on
the mode 3. Here we consider the case that the mode 3 propagates in a noisy
channel such as photon loss, and thermal noise after Alice's measurement but
before its reaching Bob's location (see Fig. 1(a)). In the interaction picture
and the Born-Markov approximation, the time evolution of the density matrix
describing the thermal environment is governed by the master equation (ME)
\cite{28}:%
\begin{align}
\frac{d}{dt}\rho \left(  t\right)   &  =\kappa \bar{n}\left(  2a^{\dag}\rho
a-aa^{\dag}\rho-\rho aa^{\dag}\right)  \nonumber \\
&  +\kappa \left(  \bar{n}+1\right)  \left(  2a\rho a^{\dag}-a^{\dag}a\rho-\rho
a^{\dag}a\right)  ,\label{14}%
\end{align}
where $\kappa$ and $\bar{n}$ are the dissipative coefficient and the average
thermal photon number of the environment, respectively. When $\bar{n}=0,$ Eq.
(\ref{14}) reduces to the one describing the photon-loss channel. By solving
the ME in the CF form, one can find that the evolution of CF described by Eq.
(\ref{14}) is given by%
\begin{equation}
\chi \left(  \gamma;t\right)  =\chi \left(  \gamma e^{-\kappa t};0\right)
\exp \left \{  -\Gamma \left \vert \gamma \right \vert ^{2}\right \}  ,\label{15}%
\end{equation}
where $\Gamma=\left(  2\bar{n}+1\right)  \left(  1-e^{-2\kappa t}\right)  /2$,
and $\chi \left(  \gamma;0\right)  $ is the CF of initial state $\rho \left(
0\right)  $. In a similar way to deriving Eq. (\ref{12}), at Bob's location,
the CF $\bar{\chi}_{f}$ of final output state for the teleportation scheme can
be directly given by%
\begin{align}
\bar{\chi}_{f}\left(  \beta;t\right)   &  =e^{-\Gamma \left \vert \beta
\right \vert ^{2}}\chi_{1}\left(  f_{1}\beta-f_{2}\beta^{\ast}\right)
\nonumber \\
&  \times \chi_{23}\left(  \beta^{\ast}f_{3}-\beta \allowbreak f_{4},\beta
e^{-\kappa t}\right)  \nonumber \\
&  \times \exp \{-R[g_{p}^{2}(\operatorname{Re}\beta)^{2}+g_{q}^{2}%
(\operatorname{Im}\beta)^{2}]\}.\label{16}%
\end{align}
The form of Eq. (\ref{16}) shows the different roles played by the noise
channel ($\Gamma,\kappa$) and gain factors ($g_{p},g_{q}$) as well as
unbalanced BS ($\theta$), the reflectivity $R$. The decoherence effect from
the noisy channel affects only mode 3 by means of the exponentially decreasing
weight $e^{-\kappa t}$ in the arguments of $\chi_{23}$. Eq. (\ref{16}) is just
the general description of the nonideal scheme in terms of the CF, which just
reduces to the factorized form of the output CF in Eq. (\ref{8}) and Eq. (4)
in Ref. \cite{17}, as expected, when $\kappa t=0$ and $g_{q}=g_{p}=g$,
$\theta=\pi/4$, respectively. Thus, Eq. (\ref{16}) is the generalized
input-output relation in the CF formalism.

\subsection{Fidelity and average fidelity}

In order to measure the effectivity of the teleportation scheme, we appeal to
the fidelity of teleportation, defined by $\mathcal{F=}$Tr$\left(  \rho
_{in}\rho_{out}\right)  $. Within the formalism of CF, the fidelity reads
\begin{equation}
\mathcal{F=}\int \frac{d^{2}\lambda}{\pi}\chi_{in}(\lambda)\chi_{out}%
(-\lambda), \label{17}%
\end{equation}
where $\chi_{in}$ and $\chi_{out}$ are the CFs corresponding to density
operators $\rho_{in}$ and $\rho_{out}$, respectively. Eq. (\ref{17}) is the
fundamental quantity that measures the performance of a CV teleportation,
which will be often used in the following calculations. On the basis of Eqs.
(\ref{16}) and (\ref{17}), we can examine the performance of teleportation for
any input states and any entangled resources including non-Gaussian ones.

In particular, when we specify the input teleported states at Alice's location
to be coherent states $\rho_{1}=\left \vert \epsilon \right \rangle \left \langle
\epsilon \right \vert $ with complex amplitude $\epsilon$, whose CF reads
$\chi_{1}\left(  \lambda \right)  =e^{-\frac{1}{2}\left \vert \lambda \right \vert
^{2}+\lambda \epsilon^{\ast}-\epsilon \lambda^{\ast}}$, then substituting it and
Eq. (\ref{16}) into Eq. (\ref{17}) we can get
\begin{align}
\mathcal{F}  &  =\int \frac{d^{2}\lambda}{\pi}\exp \left \{  -\frac{1}{2}\left(
1+f_{1}^{2}+f_{2}^{2}+2\Gamma \right)  \left \vert \lambda \right \vert
^{2}\right \} \nonumber \\
&  \times \exp \left \{  \frac{1}{2}f_{1}f_{2}\left(  \lambda^{2}+\lambda^{\ast
}{}^{2}\right)  +\lambda \Delta-\lambda^{\ast}\Delta^{\ast}\right \} \nonumber \\
&  \times \chi_{23}\left(  \lambda \allowbreak f_{4}-\lambda^{\ast}%
f_{3},-\lambda e^{-\kappa t}\right) \nonumber \\
&  \times \exp[-R(g_{p}^{2}\operatorname{Re}^{2}\lambda+g_{q}^{2}%
\operatorname{Im}^{2}\lambda)], \label{18}%
\end{align}
where $\Delta=\allowbreak \left(  1-f_{1}\right)  \epsilon^{\ast}-\epsilon
f_{2}\allowbreak$. From Eq. (\ref{18}) we can see that if we choose $\Delta=0$
then the fidelity will be independent of $\epsilon$ for any entangled
resources. The condition of $\Delta=0$ leads to%
\begin{equation}
g_{q}=\frac{1}{\sqrt{2T}\cos \theta},g_{p}=\frac{1}{\sqrt{2T}\sin \theta}.
\label{19}%
\end{equation}
This is the only choice making the fidelity independent of $\epsilon$, which
allows us to have no information about the input coherent states. The
condition (\ref{19}) depends on $T$ and $\theta$ but is independent of the
decoherence involved in mode 3. This is true for any entangled resources. In
particular, when $\theta=\pi/4$, Eq. (\ref{19}) just reduces to the case in
Ref. \cite{17}; while for $T=1$ and $\theta=\pi/4,$ this result is just the
case discussed in Ref. \cite{28a}.

Generally speaking, the fidelity in Eq. (\ref{18}) depends on the teleported
input states which are usually unknown by the sender and the receiver. Here,
in order to further describe the fidelity, we assume a partial knowledge of
the input states about a probability distribution $P\left(  \mu \right)  $
satisfying the normalization condition, i.e., $\int P\left(  \mu \right)
d\mu=1$ where the integral is taken over\ all possible values of $\mu$. For a
given distribution $P\left(  \mu \right)  $, the average fidelity is
\begin{equation}
\bar{F}=\int \mathcal{F}\left(  \mu \right)  P\left(  \mu \right)  d\mu.
\label{20}%
\end{equation}
In the following, we will take three probability distributions into account
for input coherent states, such as line-, circle- and 2D-Gaussian-distribution
\cite{19}.

\section{Two-mode squeezed vacuum as entangled resources}

In this section, we use the usual TMSV as entangled resources to analyze the
performance of these three tunable parameters for improving the fidelity of
teleportation. The TMSV entangled resource, most commonly used in
continuous-variable teleportation, can be generated by the parametric
down-conversion (PDC) process, and theoretically can be defined as%
\begin{align}
\left \vert \Phi \right \rangle _{sv}  &  =S\left(  r\right)  \left \vert
00\right \rangle \nonumber \\
&  =\text{sech}r\exp \left(  a^{\dagger}b^{\dagger}\tanh r\right)  \left \vert
00\right \rangle , \label{21}%
\end{align}
where $S\left(  r\right)  =\exp \left \{  r(a^{\dagger}b^{\dagger}-ab)\right \}
$ is the two-mode squeezing opertor with $r$ being the squeezing parameter,
and $a^{\dagger}\left(  a\right)  $ and $b^{\dagger}\left(  b\right)  $ are
photon creation (annihilation) operators. According to the definition of CF,
the CF of the TMSV is given by
\begin{align}
\chi_{sv}\left(  \alpha,\beta \right)   &  =\exp \left \{  -\frac{1}{2}\left(
\left \vert \alpha \right \vert ^{2}+\left \vert \beta \right \vert ^{2}\right)
\cosh2r\allowbreak \right \} \nonumber \\
&  \times \exp \left \{  \frac{1}{2}\left(  \alpha \beta+\alpha^{\ast}%
\allowbreak \beta^{\ast}\right)  \sinh2r\right \}  . \label{22}%
\end{align}
In particular, for the largest entangled resource with $r\rightarrow \infty$,
and ideal measurements with $T=1$ and $R=0$, as well as banlanced BS, we have
$\lim_{r\rightarrow \infty}\chi_{sv}\left(  \beta^{\ast},\beta \right)  =1$.
Substituting it into Eq. (\ref{12}) yields $\lim_{r\rightarrow \infty}\bar
{\chi}_{out}\left(  \beta \right)  =\chi_{1}\left(  \beta \right)  $, i.e., a
perfect teleportation, as expected.

When Alice use the TMSV as entangled resource to teleport the coherent states,
then the fidelity in Eq. (\ref{18}) can be calculated as%
\begin{equation}
\mathcal{F}=\frac{1}{\sqrt{G}}\exp \left \{  \frac{-K_{1}\left \vert
\Delta \right \vert ^{2}+K_{2}\left(  \Delta^{2}+\Delta^{\ast2}\right)  }%
{G}\right \}  , \label{22a}%
\end{equation}
where we have defined $G=K_{1}^{2}-4K_{2}^{2}$ and%
\begin{align}
K_{1}  &  =\frac{1}{2}\left(  1+f_{1}^{2}+f_{2}^{2}+2\Gamma \right)  +\frac
{R}{2}\left(  g_{p}^{2}+g_{q}^{2}\right) \nonumber \\
&  +\frac{1}{2}\left(  f_{3}^{2}+f_{4}^{2}+e^{-2\kappa t}\right)
\cosh2r-f_{3}e^{-\kappa t}\sinh2r,\nonumber \\
K_{2}  &  =\frac{1}{2}\left \{  f_{1}f_{2}-R\left(  g_{p}^{2}-g_{q}^{2}\right)
/2\right. \nonumber \\
&  +\left.  f_{3}f_{4}\cosh2r-f_{4}e^{-\kappa t}\sinh2r\right \}  , \label{22b}%
\end{align}
and used the following integration formula%
\begin{align}
&  \int \frac{d^{2}z}{\pi}\exp \left(  \zeta \left \vert z\right \vert ^{2}+\xi
z+\eta z^{\ast}+fz^{2}+gz^{\ast2}\right) \nonumber \\
&  =\frac{1}{\sqrt{\zeta^{2}-4fg}}\exp \left \{  \frac{-\zeta \xi \eta+\xi
^{2}g+\eta^{2}f}{\zeta^{2}-4fg}\right \}  . \label{22c}%
\end{align}
From Eq. (\ref{22a}) one can see that the fidelity $\mathcal{F}$ depends on
the amplitude of the teleported coherent states. In the next sections, we
shall consider two kinds of cases: one is independent of the amplitude by the
choice in Eq. (\ref{19}) and the other is not, but partial information about
the input state distribution is known.

\section{$\epsilon$-independent optimal fidelity}

In this section, we examine the fidelity for teleporting coherent state with
two gain factors fixed to be $g_{q}=1/(\sqrt{2T}\cos \theta)$, $g_{p}%
=1/(\sqrt{2T}\sin \theta)$. This choice allows us to have no information about
the amplitude of coherent states. Noticing that $f_{1}=1$, $f_{2}=0$,
$f_{3}=\csc2\theta$ and $f_{4}=-\cot2\theta$, then from Eq. (\ref{22a}) we can
get%
\begin{equation}
\mathcal{F}_{\epsilon}=\left \{  H[1/(\sqrt{2T}c_{1}),c_{2}]H[1/(\sqrt{2T}%
c_{2}),c_{1}]\right \}  ^{-1/2}, \label{23}%
\end{equation}
where $c_{1}=\cos \theta,c_{2}=\sin \theta$ and we defined the function
$H\left(  x,y\right)  $ as
\begin{align}
H\left(  x,y\right)   &  =\frac{1}{2}+\Gamma+x^{2}\left(  1+2Ty^{2}\sinh
^{2}r\right) \nonumber \\
&  +\frac{1}{2}e^{-2\kappa t}\cosh2r-xye^{-\kappa t}\sqrt{2T}\sinh2r.
\label{33}%
\end{align}
It is clear that the fidelity $\mathcal{F}_{\epsilon}$ depends on
multi-parameters such as $r,$ $\kappa t,$ $\bar{n},$ $T$ and $\theta$. At
fixed $r,$ $\kappa t,$ $\bar{n}$ and $T$, the optimal fidelity of
teleportation is defined as%
\begin{equation}
\mathcal{F}_{opt}=\max_{\theta}\mathcal{F}\left(  r,\theta \right)  .
\label{25}%
\end{equation}
In order to maximize the fidelity in Eq. (\ref{23}) over $\theta$, we can take
$\partial \mathcal{F}_{\epsilon}/\partial \theta=0$, which leads to the
following condition%
\begin{equation}
\cos2\theta=0, \label{26}%
\end{equation}
or%
\begin{align}
\csc2\theta &  =\frac{1}{2}\left \{  \frac{e^{-\kappa t}\sinh2r}{1/T+2\sinh
^{2}r}\right.  +\nonumber \\
&  \left.  \frac{e^{-\kappa t}\sinh2r}{1/T+1+2\Gamma+e^{-2\kappa t}\cosh
2r}\right \}  . \label{27}%
\end{align}
It is easy to see that the first item (FI) in the right hand side (RHS) of Eq.
(\ref{27}) is less than unit, and the second item (SI) satisfies (by taking
$T=1$ and $\bar{n}=0$)%
\begin{equation}
\frac{e^{-\kappa t}\sinh2r}{1/T+1+2\Gamma+e^{-2\kappa t}\cosh2r}\leqslant
\frac{e^{-\kappa t}\sinh2r}{3+2e^{-2\kappa t}\sinh^{2}r}. \label{28}%
\end{equation}
By numerical calculation, we can find that when the squeezing parameter $r$ is
less than a threshold value of about $2.1$, the sum of $($FI$+$SI$)/2$ is
alway less than unit which will lead to an impossible case, i.e., $\csc
2\theta<1$. Thus within the region of threshold value, the optimal point is at
$\theta=\pi/4$ and $g_{q}=g_{p}=1/\sqrt{T}$ which is independent of $\bar{n}$
and $e^{-\kappa t}$. The threshold value of $r$ will increase with the
increasing $\bar{n}$ and $1/T$. Actually, the presence of threshold value
results from the decoherence through mode 3, since the SI is always less than
unit for any squeezing $r$ when $\kappa t=0$.

Substituting the above optimal condition into Eq. (\ref{23}), we get the
optimal fidelity%
\begin{equation}
\mathcal{F}_{opt}=[\frac{1}{T}+\Gamma+e^{-\kappa t}(\cosh \kappa t\cosh
2r-\sinh2r)]\allowbreak^{-1}. \label{29}%
\end{equation}
It is clear that $\mathcal{F}_{opt}$ decreases with the increasing $\bar{n}$
and $1/T$, as expected. In particular, when $\kappa t=0$ and $T=1$, Eq.
(\ref{29}) just reduces to $\mathcal{F}_{opt}=(1+\tanh r)/2$, which is the
best fidelity when we use the coherent states as inputs and the TMSV as
entangled resources in the BK scheme independent of teleported coherent state
amplitude. In addition, when $\kappa t\rightarrow \infty$, $\mathcal{F}%
_{opt}\rightarrow(\frac{1}{T}+\bar{n}+\cosh^{2}r)^{-1}$, which decreases with
the increasing $\bar{n},r$ and the decreasing $T$.

\begin{figure}[ptb]
\label{Fig2}
\centering \includegraphics[width=0.9\columnwidth]{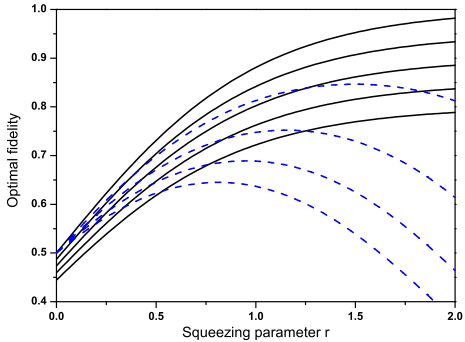}\caption{(Colour
online) The optimal fidelity for teleporting coherent states as a function of
the squeezing parameter $r$ with some different values of $T$ and $\kappa t$
as well as $\bar{n}=0$. Here solid lines: the transmissivity
$T=1,0.95,0.9,0.85,0.8$ and $\kappa t=0;$ Blue-dash lines: $\kappa
t=0.1,0.2,0.3,0.4$ and $T=1$. The corresponding lines are arranged from top to
bottom with the increasing $1/T$ and $\kappa t$, respectively.}%
\end{figure}

In order to examine clearly the effects of different parameters on the optimal
fidelity, we plot the optimal fidelities as a function of squeezing parameter
$r$ in Fig. 2. From Fig. 2, we can see that for $\kappa t=0$ (without
decoherence on mode 3) the optimal fidelities increase monotonically with the
increasing squeezing parameter $r$ and the transmissivity $T$. In addition,
when we consider the effects of decoherence on mode 3, the optimal fidelities
first increase and then decrease with the increasing $r$. The maximal value
and the corresponding value of $r_{\max}$ reduces as $\kappa t$ increases. In
fact, we can take $\partial \mathcal{F}_{opt}/\partial r=0$ to get a simple
relation as following ($\cosh \kappa t=\coth2r_{\max})$%
\begin{equation}
e^{2r_{\max}}=\coth \frac{\kappa t}{2}. \label{30}%
\end{equation}
It is interesting to notice that $r_{\max}$ is independent of $T$ and $\bar
{n}$.

\section{Average optimal fidelity and the effect of tunable parameters}

In the last section, we consider the $\epsilon$-independent optimal fidelity.
However, when $g_{q}\neq1/(\sqrt{2T}\cos \theta)$ and $g_{p}\neq1/(\sqrt
{2T}\sin \theta)$, the scenario will be changed dramatically. In this case, the
fidelity in (\ref{22a}) will depend on the amplitude $\epsilon$ of coherent
state. In this section, we examine the average optimal fidelity for three
different probability distributions of the teleported input states in which
the partial information is known by Alice and Bob. For example, they may be
completely sure of phase of the states but the amplitude is unknown \cite{19}.

\subsection{Optimal fidelity for teleporting coherent states on a line}

First, let us consider the teleportation of coherent states on a line. Without
losing the generality, here we assume that the phase of the teleported
coherent states is zero because we can always achieve this point by rotating
frame. Then the corresponding probability distribution can be given by
(letting $\epsilon=x+iy$)%
\begin{equation}
P\left(  x,y\right)  =\frac{1}{2L}\delta \left(  y\right)  \times \left \{
\begin{array}
[c]{cc}%
1, & \left \vert x\right \vert \leqslant L\\
0, & \text{else}%
\end{array}
\right.  . \label{31}%
\end{equation}
Thus substituting Eqs. (\ref{31}) and (\ref{22a}) into Eq. (\ref{20}) yields
the average fidelity%
\begin{align}
\mathcal{\bar{F}}_{line}  &  =\frac{1}{2L\sqrt{G}}\int_{-L}^{L}dxe^{-\frac
{M}{G}x^{2}}\nonumber \\
&  =\frac{\sqrt{\pi}}{2L\sqrt{M}}\text{Erf}\left \{  \frac{|1-\sqrt{2T}%
g_{q}\cos \theta|}{[H\left(  g_{q},\sin \theta \right)  ]^{1/2}L^{-1}}\right \}  ,
\label{32}%
\end{align}
where Erf$\{a\}=$ $1/\sqrt{\pi}\int_{-a}^{a}e^{-x^{2}}dx$ is the error
function and $M\left.  =\right.  (1-\sqrt{2T}g_{q}\cos \theta)^{2}H\left(
g_{p},\cos \theta \right)  $.

Noticing the separability of $g_{q}$ and $g_{p}$ in $\mathcal{\bar{F}}_{line}%
$, the optimal value of $g_{p}$ can be obtained by $\partial \mathcal{\bar{F}%
}_{line}/\partial g_{p}=0$ equivalent to $\partial H\left(  g_{p},\cos
\theta \right)  /\partial g_{p}=0,$ which leads to
\begin{equation}
g_{p}^{opt}=\frac{e^{-\kappa t}\sqrt{2T}\cos \theta^{opt}\sinh2r}%
{2(1+2T\cos^{2}\theta^{opt}\sinh^{2}r)}. \label{34}%
\end{equation}
It is interesting to notice that the optimal value of $g_{p}^{opt}$ is related
to $e^{-\kappa t}$ and $T$ but independent of the average thermal photon-number.

Next, we will maximize the fidelity by numerical calculation. At fixed $r$,
$\kappa t$, $\bar{n}$ and $T$, the optimal fidelity of teleportation can be
defined as
\begin{equation}
\mathcal{\bar{F}}_{line}^{opt}=\max_{g_{q},g_{p},\theta}\mathcal{\bar{F}%
}_{line}\left(  r,g_{q},g_{p},\theta \right)  . \label{35}%
\end{equation}
In Fig. 3 we plot the optimal fidelity as a function of squeezing parameter
$r$ for some different values of parameters $\kappa t$, $\bar{n}$ and $T$. In
Fig. 3(a), we consider the optimal fidelities with some different values of
$L$ and $T=1,\bar{n}=0$ as well as $\kappa t=0.2$ (for comparison, the case of
$\kappa t=0$ is also plotted as dash lines). From Fig. 3(a), we can see that
the optimal fidelities grow with increasing $r$ and $1/L$. The fidelities can
be greatly optimized with respect to the standard teleportation scheme lines
(STS with $g_{q}=g_{p}=1$ and $\theta=\pi/4$, see short dash-dot-(dot) lines).
Especially for a smaller $L$ ($L=0.1$), the optimal fidelity can almost access
to unit. While for a larger $L$ (say $L=300$), the fidelity achieves a
limitation (still over 0.8) which is still superior to that in the STS. In
Fig. 3 (b), we consider the effect of different values of $T$ on the optimal
fidelity at $\kappa t=0,0.2$. It is shown that the optimal values decease with
the increasing $1/T$ for a given $\kappa t$; while for the case of $T\neq1$,
by comparing the fidelities at $\kappa t=0,0.2$ for a given $T$,\ it is found
that the optimal fidelities first increase and then decrease with the
increasing $r$, and it is interesting to notice that the optimal fidelity with
$\kappa t=0.2$ is superior to that with $\kappa t=0$ when $r$ exceeds a
certain cross-point.

\begin{figure}[ptb]
\label{Fig3} \centering \includegraphics[width=0.9\columnwidth]{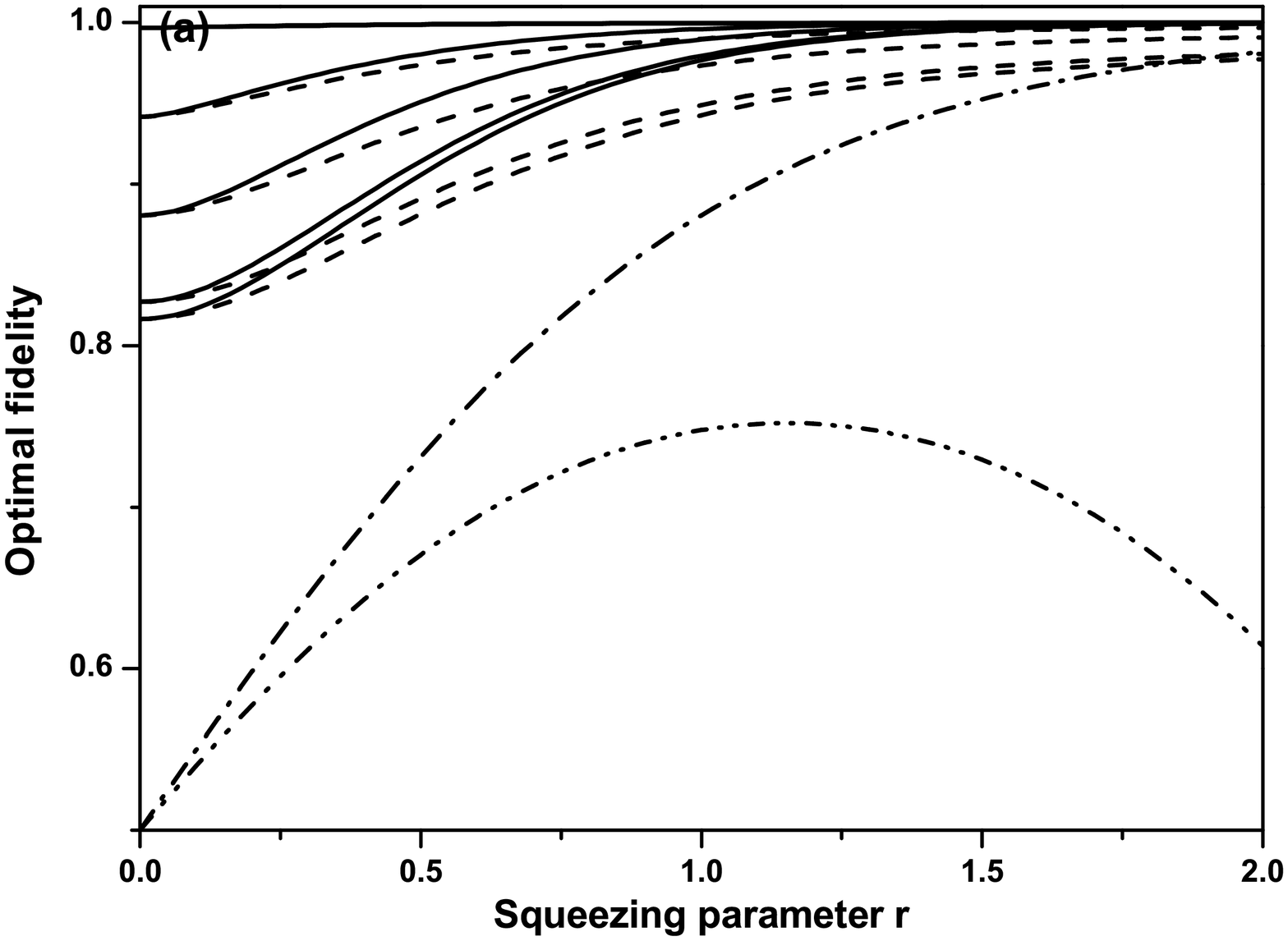}
\newline \includegraphics[width=0.9\columnwidth]{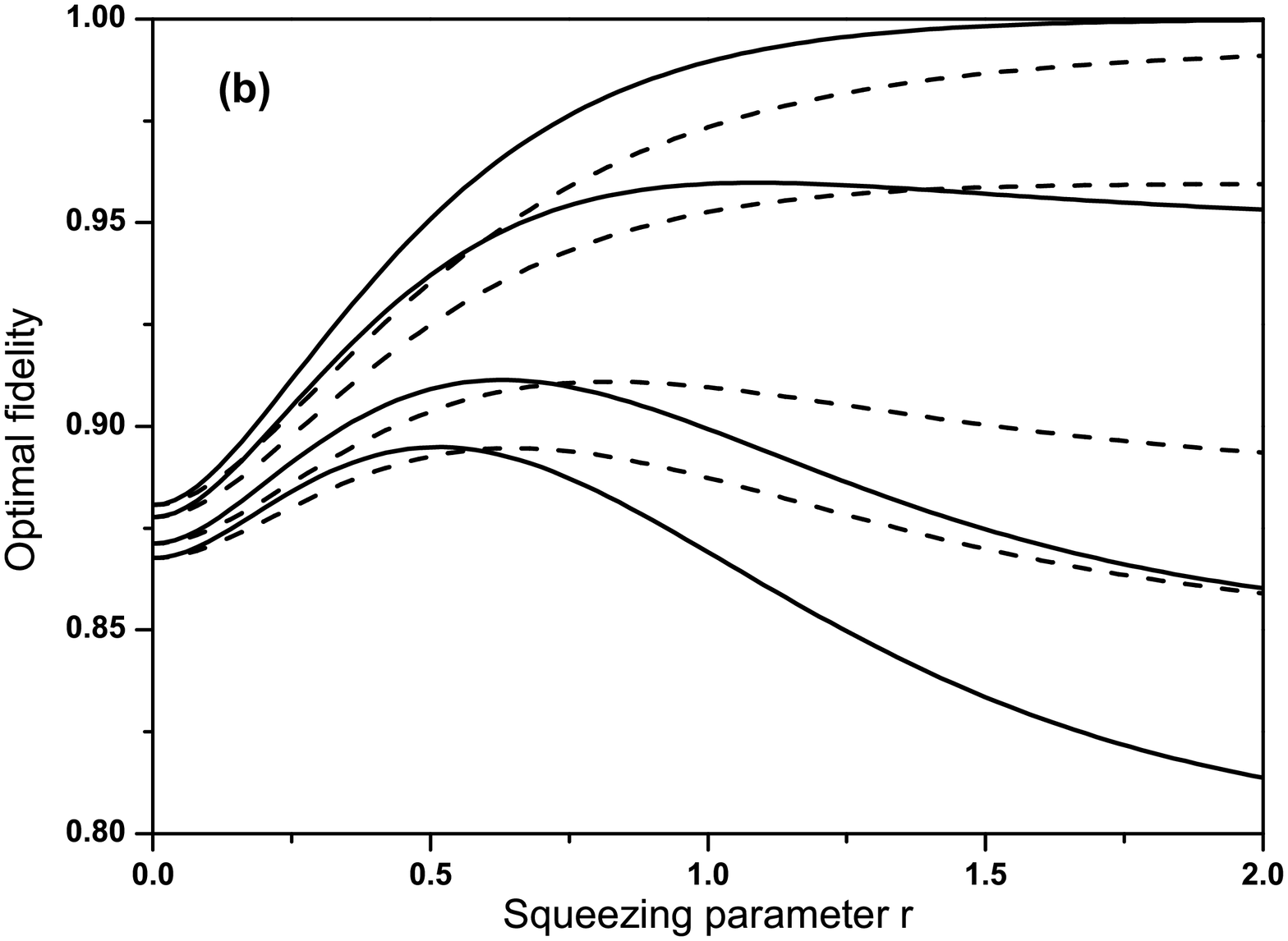}\caption{The optimal
fidelity for teleporting CSs as a function of $r$ with $\bar{n}=0$, $\kappa
t=0,0.2$ corresponding to solid and dash lines, respectively. (a)
$L=0.1,0.5,1,3,300$ and $T=1;$ for comparison, the teleportation in the STS is
also plotted as dash-dot and dash-dot-dot lines for $\kappa t=0,0.2$; (b)
$T=1,0.95,0.9,0.85,0.8$ and $L=1$. The corresponding lines are arranged from
top to bottom with the increasing $L$ and $1/T$ for a given $\kappa t$.}%
\end{figure}

Furthermore, in order to clearly see whether there is the ability against the
decoherence by using these tunable parameters, here we plot the optimied
fidelity as a function of the evolution time $\kappa t$ for some fixed values.
For this purpose, we fix the squeezing parameter at the intermediate value
$r=0.8$ with $T=0.9$, $\bar{n}=0$. Experimentally, the attainable squeezing
degree is about $1.5$. Fig. 4 shows that the optimal fidelity remains above
0.8 which exceeds the classical threshold up to significantly large values of
$\kappa t$. This case is true even for the limitation of $L\rightarrow \infty$.
These results indicate that the optimizal fidelities by three parameters
present superior behavior to and higher ability against the decoherence than
those in the standard teleportation scheme (with $g_{q}=g_{p}=1$ and
$\theta=\pi/4$, also see dash lines in Fig. 4).

\begin{figure}[ptb]
\label{Fig4}
\centering \includegraphics[width=0.9\columnwidth]{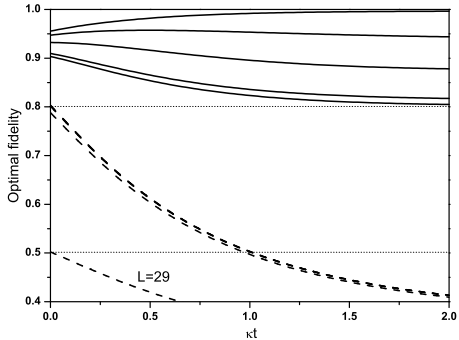}\caption{(Colour
online) The optimal fidelity for teleporting CSs as a function of $\kappa t$
with $\bar{n}=0$, $r=0.8,$ $T=0.9$ and $L=0.1,0.5,1,3,300$ from top to bottom,
respectively.}%
\end{figure}

In addition, we make a comparison among the optimal effects of different
optimal parameters. In Fig. 5, we plot the optimized fidelity over different
tunable parameters as the function of squeezing parameter $r$ for a given
$L=1$ with $T=1$, $\bar{n}=0$ and $\kappa t=0.2$. It is found that the
optimization by three tunable parameters is the best when compared to single-
and two-parameter optimization, especially in the region of small
entanglement, which indicates that the role of parameter is different from
each other. Thus, it is necessary to perform a simultaneous balanced
optimization over these three parameters to obtain a maximization of
teleportation fidelity for the probability distribution in Eq. (\ref{31}).

\begin{figure}[ptb]
\label{Fig5}
\centering \includegraphics[width=0.9\columnwidth]{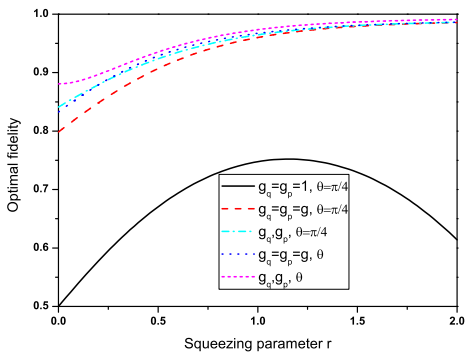}\caption{(Colour
online) The optimal fidelity for teleporting CSs as a function of $r$ with
$\bar{n}=0$, $\kappa t=0.2,$ $T=L=1$ for several different optimal
parameters.}%
\end{figure}

\subsection{Optimal fidelity for teleporting CSs on a circle}

In this subsection, we consider the optimal fidelity for teleporting CSs on a
circle, which means $\epsilon=\left \vert \epsilon \right \vert e^{i\varphi
}\equiv Ae^{i\varphi}$ with a known amplitude $\left \vert \epsilon \right \vert
=A$ and unknown phase $\varphi$. In this case, the distribution function is
$P\left(  A,\varphi \right)  =\delta \left(  \left \vert \epsilon \right \vert
-A\right)  /2\pi$ satisfying $\int_{0}^{\infty}\int_{0}^{2\pi}P\left(
A,\varphi \right)  d\left \vert \epsilon \right \vert d\varphi=1$, then the
average fidelity can be calculated as%
\begin{equation}
\mathcal{\bar{F}}_{circle}=\frac{e^{-R_{1}}}{\sqrt{G}}\sum_{k=0}^{\infty}%
\frac{\left(  R_{2}\right)  ^{2k}}{k!k!}, \label{36}%
\end{equation}
where we have set $R_{1}=A^{2}\{K_{1}[\left(  1-f_{1}\right)  ^{2}+f_{2}%
^{2}]+4K_{2}\left(  1-f_{1}\right)  f_{2}\}/G$, $R_{2}=A^{2}\{K_{1}\left(
1-f_{1}\right)  f_{2}+K_{2}[\left(  1-f_{1}\right)  ^{2}+f_{2}^{2}]\}/G$.
Maximizing $\mathcal{\bar{F}}_{circle}$ over these three parameters, we can
get the optimized fidelity $\mathcal{\bar{F}}_{circle}^{opt}=\max_{g_{q}%
,g_{p},\theta}\mathcal{\bar{F}}_{circle}\left(  r,g_{q},g_{p},\theta \right)
$. Our random numerical calculations show that, for the probability
distribution, the maximum value of fidelity can be achieved at the point with
$g_{q}=g_{p}=g$ and $\theta=\pi/4$, which is different from the case in
subsection A. Under this condition we have $f_{1}=f_{3}=g\sqrt{T}$ and
$f_{2}=f_{4}=0$, as well as $R_{2}=K_{2}=0$. Thus the optimized fidelity can
be given by%
\begin{equation}
\mathcal{\bar{F}}_{circle}^{opt}=\frac{1}{\Theta}\exp \left \{  -\frac{A^{2}%
}{\Theta}(1-g\sqrt{T})^{2}\right \}  , \label{37}%
\end{equation}
where we have set $\Theta=\Gamma+[\allowbreak g^{2}(R+1)+1+(g\sqrt
{T}-e^{-\kappa t})^{2}\cosh2r+2g\sqrt{T}e^{-\kappa t}e^{-2r}]/2$. It is
obvious that $\Theta>0$.

Using Eq. (\ref{36}) or (\ref{37}), we have plotted the optimal fidelity as a
function of squeezing parameter $r$ for some different values of $A$ and $T$
in Fig. 6. In Fig. 6(a), we consider the optimal fidelities with some
different values of $A$ with $T=1,\bar{n}=0$ as well as $\kappa t=0,0.2$. From
Fig. 6(a), we can see that the optimal fidelities grow monotonously with
increasing $r$ for $\kappa t=0,$ but for $\kappa t=0.2$ the optimal fidelities
first increase and then decrease with increasing $r$ especially for a large
$A$ (say $A=3$). In addition, for a small $A$, the optimal fidelity almost
access to unit. In Fig. 6(b), we also examine the effect of different $T$ on
the fidelity. It is interesting to notice that the optimal fidelity with
$\kappa t=0.2$ can be better than that with $\kappa t=0$ when the squeezing
$r$ exceeds a certain value. This case is similar to that in Fig. 3(b). In
Fig. 6, the point $r_{\max}$ corresponding to the maximum fidelity depends on
$\kappa t,$ and for given $\kappa t$, $A$ and $T$, the value of $r_{\max}$ can
be determined by taking $\partial \mathcal{\bar{F}}_{circle}^{opt}/\partial
r=0,$ which leads to
\begin{equation}
\left \{  \Theta-A^{2}(1-g\sqrt{T})^{2}\right \}  \frac{\partial \Theta}{\partial
r}=0. \label{38}%
\end{equation}
After a straightforward calculation, we can abtain%
\begin{equation}
\tanh2r_{\max}=\frac{2g\sqrt{T}e^{-\kappa t}}{g^{2}T+e^{-2\kappa t}}.
\label{39}%
\end{equation}
or%
\begin{equation}
e^{4r_{\max}}-be^{2r_{\max}}+c^{2}=0, \label{40}%
\end{equation}
where $b=4[A^{2}(1-g\sqrt{T})^{2}-\Gamma]/(g\sqrt{T}-e^{-\kappa t})^{2},$ and
$c=(g\sqrt{T}+e^{-\kappa t})/(g\sqrt{T}-e^{-\kappa t})$. In particular, when
$g=1/\sqrt{T}$ the fidelity is independent of the amplitude $A$ and since that
$\Theta>0$, the value of $r_{\max}$ in Eq. (\ref{39}) reduces to that in Eq.
(\ref{30}).

\begin{figure}[ptb]
\label{Fig6} \centering \includegraphics[width=0.9\columnwidth]{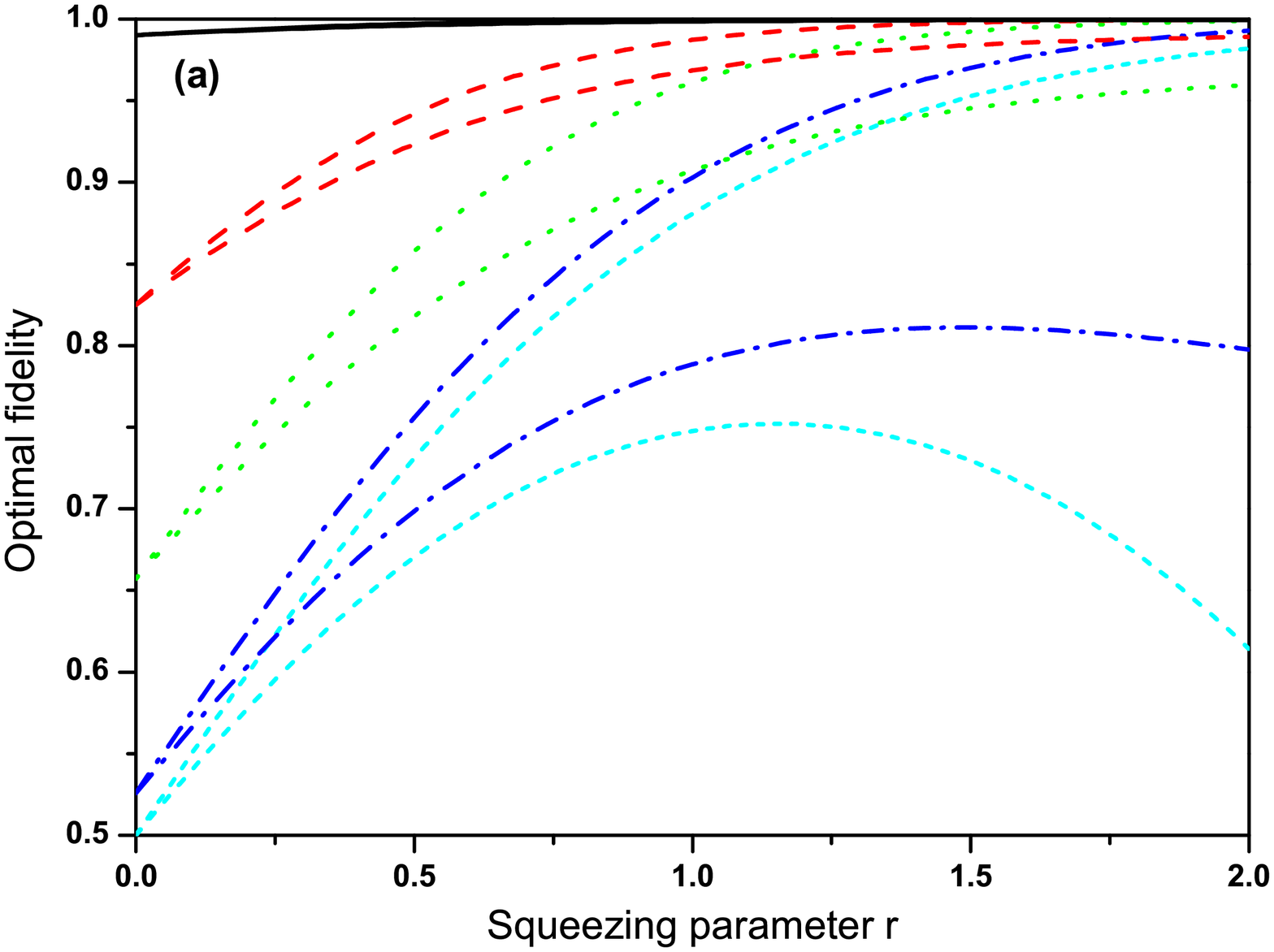}
\newline \includegraphics[width=0.9\columnwidth]{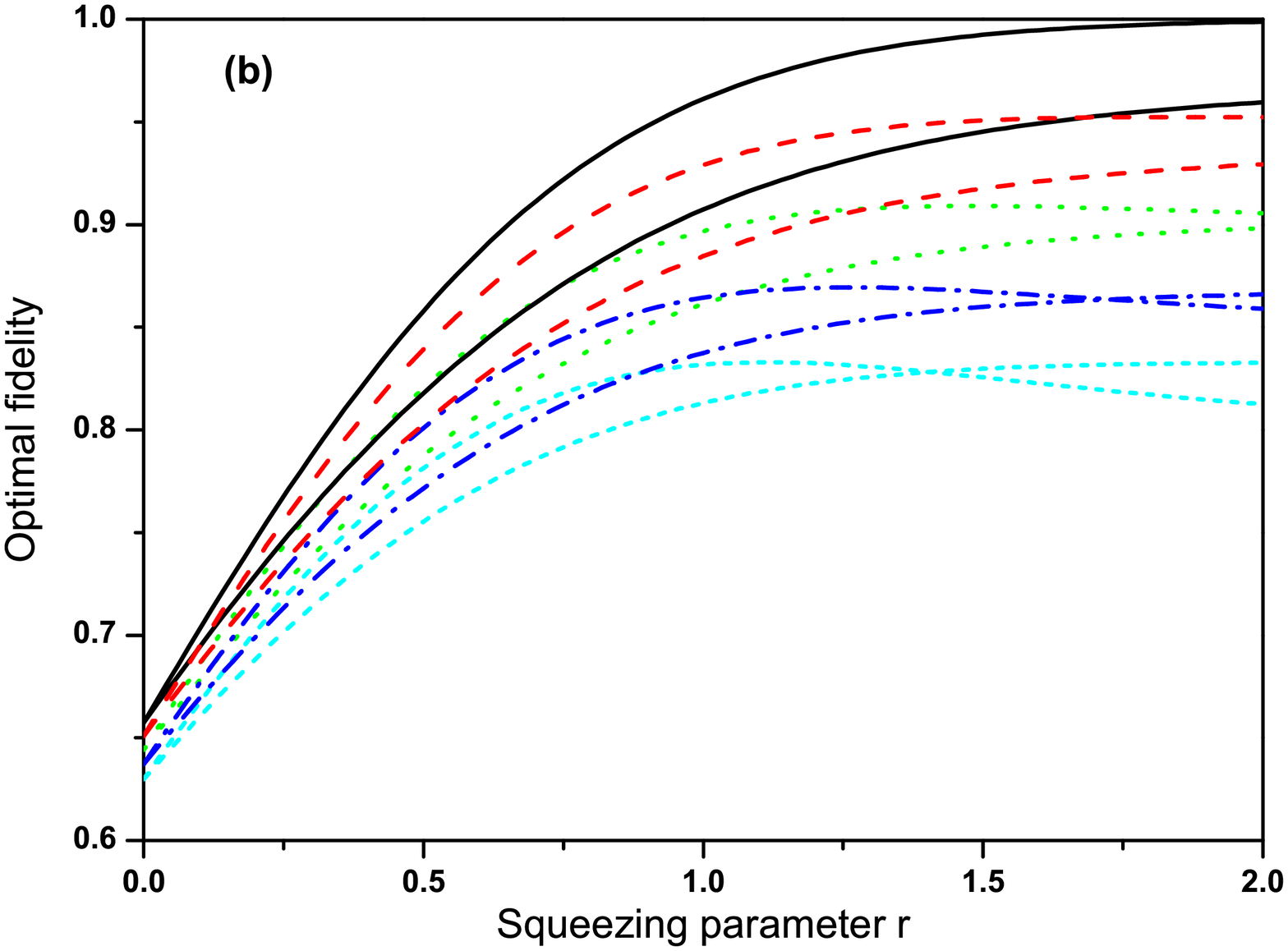}\caption{(Colour
online) The optimal fidelity for teleporting CSs as a function of $r$ with
$\bar{n}=0$, $\kappa t=0,0.2$ (a) $A=0.1,0.5,1,3,300$ and $T=1;$ (b)
$T=1,0.95,0.9,0.85,0.8$ and $A=1$. For each optimized case (associated with a
special plot style), the corresponding lines are arranged from top to bottom
with the increasing $A$ and $1/T$ at the point $r=0$, respectively.}%
\end{figure}In Fig. 7, choosing the same values of parameters $T,$ $r$ and
$\bar{n}$ as in Fig. 4, we plot the optimal fidelity as a function of $\kappa
t$. For comparison, we also plot the fidelity without the optimization, i.e.,
$g_{q}=g_{p}=1$ and $\theta=\pi/4$ (see dash lines). Fig. 7 shows that the
teleportation fidelity can be always above the classical limitation 0.5 up to
significantly large values of $\kappa t\left(  \leqslant2\right)  $\ when the
amplitude $A$ is less than about $1.7$, and while it can go below the
limitation $0.5$ for $A>1.7$ when $\kappa t$ exceeds a certain threshold
value. The optimal fidelities with $A=15,300$ are indistinguishable. In the
STS, the fidelity is less than 0.5 when $A>15$. Comparing the fidelities with
and without optimization, it is shown that the former have better
teleportation performance than the latter. However, this improvement is
inferior to that shown in Fig. 4 where the optimal fidelities are over 0.8 for
any $L$ and $\kappa t$.

\begin{figure}[ptb]
\label{Fig7}
\centering \includegraphics[width=0.9\columnwidth]{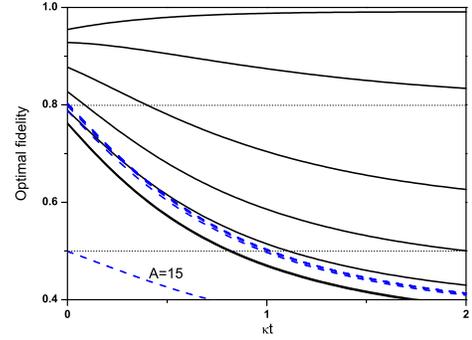}\caption{(Colour
online) The optimal fidelity for teleporting CSs as a function of $\kappa t$
with $\bar{n}=0$, $r=0.8,$ $T=0.9$ and $A=0.1,0.5,1,1.7,3,300$ from top to
bottom, respectively.}%
\end{figure}

\subsection{Optimal fidelity for teleporting CSs by 2D Gaussian distribution}

In the last subsection, we consider another simple probability
distribution-----two-dimensional (2D) Gaussian distribution. The corresponding
distribution is given by $P\left(  \alpha \right)  =1/(\pi \chi)\exp[-\left \vert
\alpha \right \vert ^{2}/\chi]$ satisfying $\int P\left(  \alpha \right)
d^{2}\alpha=1$ \cite{17,19,29}, where the variance parameter $\chi$ determines
the cutoff of the amplitude $\alpha$. Thus, using Eqs. (\ref{20}) and
(\ref{22a}), the averaged fidelity can be calculated as%
\begin{align}
\mathcal{\bar{F}}_{G}  &  =\frac{1}{\sqrt{H(g_{p},\cos \theta)+\chi(1-\sqrt
{2T}g_{p}\sin \theta)^{2}}}\nonumber \\
&  \times \frac{1}{\sqrt{H(g_{q},\sin \theta)+\chi(1-\sqrt{2T}g_{q}\cos
\theta)^{2}}}, \label{41}%
\end{align}
where the function $H(x,y)$ is defined in Eq. (\ref{33}) [$\left(
K_{1}+2K_{2}\right)  =H\left(  g_{q},\sin \theta \right)  $, $\left(
K_{1}-2K_{2}\right)  =H\left(  g_{p},\cos \theta \right)  $]. Noticing that the
parameters $g_{q}$ and $g_{p}$ are independent from each other in Eq.
(\ref{41}), thus it is not hard to obtain the optimal point by $\partial
\mathcal{\bar{F}}_{G}/\partial g_{q}=\partial \mathcal{\bar{F}}_{G}/\partial
g_{p}=\partial \mathcal{\bar{F}}_{G}/\partial \theta=0$, i.e., $\theta=\pi/4$
and $g_{q}=g_{p}=g$, where$\ g$ is given by
\begin{equation}
g=\frac{\sqrt{T}\left(  e^{-\kappa t}\sinh2r+2\chi \right)  }{2[1+T(\sinh
^{2}r+\chi)]}. \label{42}%
\end{equation}
At this optimal point, the average optimized fidelity can be expressed as
\begin{equation}
\mathcal{\bar{F}}_{G}^{opt}=\frac{1}{H(g,1/\sqrt{2})+\chi(1-g\sqrt{T})^{2}}.
\label{43}%
\end{equation}
It is clearly seen that the optimal factor $g$ depends not only on $T$, but
also on the evolution time $\kappa t$ in a different form. In particular, when
$\allowbreak T\rightarrow1$ and $\kappa t\rightarrow0$, Eq. (\ref{42}) reduces
to the result in Ref. \cite{30}. In addition, in the limitation case of
$\chi \rightarrow \infty$ which implies that the probability distribution
includes the whole complex plane, then we have $g\rightarrow1/\sqrt{T}$, which
just corresponds to the fidelity independent of $\epsilon$.

Using Eq. (\ref{41}) or (\ref{43}), we have plotted the optimal fidelity as a
function of squeezing parameter $r$ and $\kappa t$ for some different values
of $\chi$ in Fig. 8 and Fig. 9, respectively. From Fig. 8, we can see that the
smaller the distribution $\chi$ is, the higher the optimal fidelity is. As
$\chi$ increases which implies that we have less knowledge of the amplitude of
the teleported states, the optimal fidelity approaches to that in the standard
scheme ($g=1$). In addition, as $r$ increases, the fidelity first increases up
to a $\kappa t$-dependent maximum $r_{\max}$, and then decreases for a larger
values of $r$ for a given big $T\chi$. Actually, using $\partial
\mathcal{\bar{F}}_{G}^{opt}/\partial r=0$, we can get%
\begin{equation}
e^{2r_{\max}}=\frac{\left(  e^{\kappa t}+1\right)  T\chi+1}{\left(  e^{\kappa
t}-1\right)  T\chi-1}. \label{44}%
\end{equation}
For instance, when $T=1,\chi=300$ and $\kappa t=0.2$, then $r_{\max}\simeq1.\,
\allowbreak16$, which is in agreement with the numerical result in Fig. 8. In
Fig. 9, we also consider the effect of decoherence on the fidelity. We can see
the similar results to the case of circle distribution. Among these three
distributions used above, the line distribution presents the most improvement
for fidelity, but the Gaussian distribution presents the lowest improvement.
However, a common advantage is that the fidelity with CV can be improved by
using the tunable parameters even in the environments when compared with the
standard teleportation scheme.

\begin{figure}[ptb]
\label{Fig8}
\centering \includegraphics[width=0.9\columnwidth]{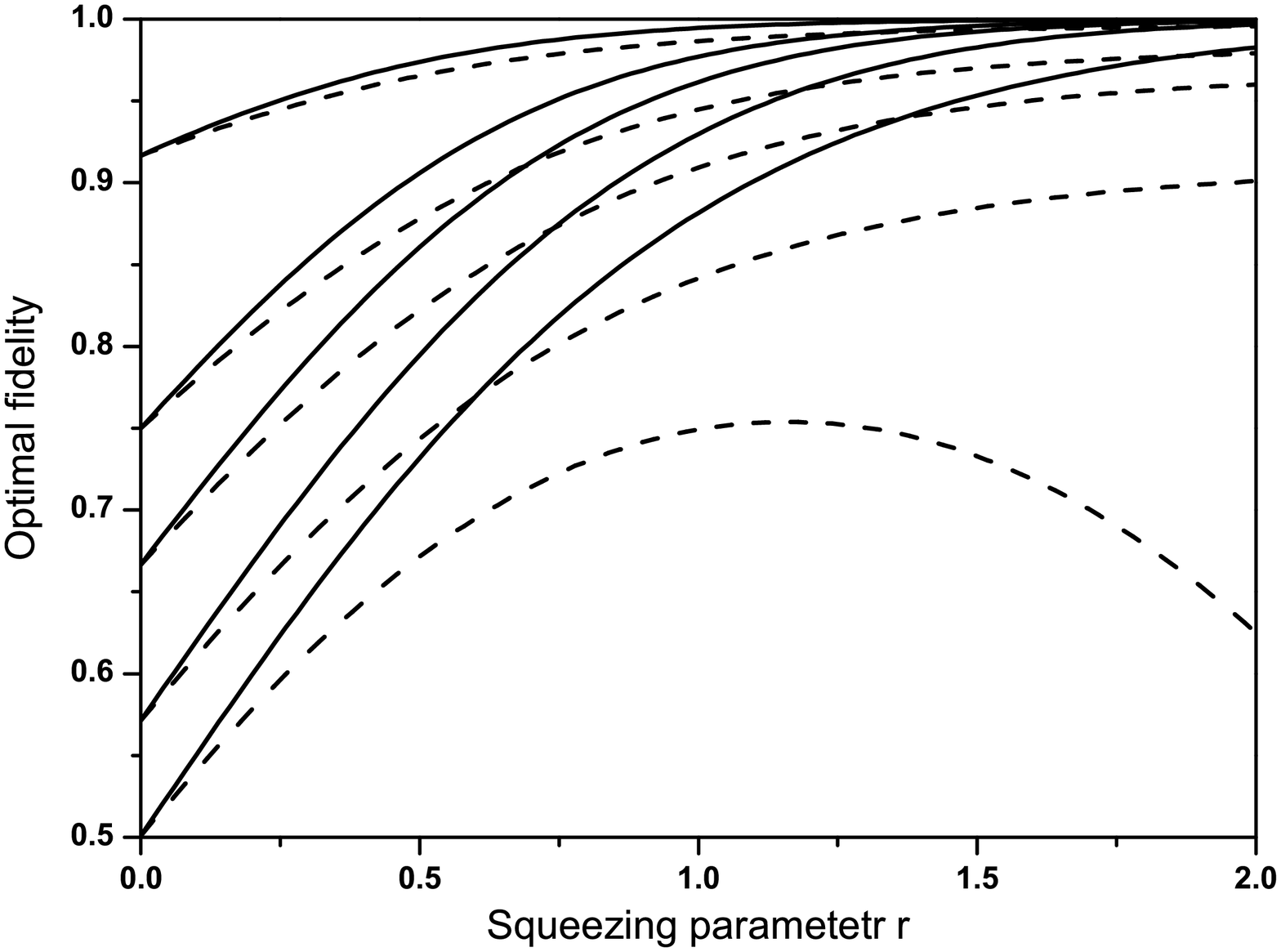}\caption{(Colour
online) The optimal fidelity for teleporting CSs as a function of $r$ with
$\bar{n}=0$, $T=1,$ $\chi=0.1,0.5,1,3,300$ and $T=1.$ For each optimized case
(associated with a special plot style), the corresponding lines are arranged
from top to bottom with the increasing $\chi$. $\kappa t=0,0.2$ correspond to
solid and dash lines, respectively. }%
\end{figure}

\begin{figure}[ptb]
\label{Fig9}
\centering \includegraphics[width=0.9\columnwidth]{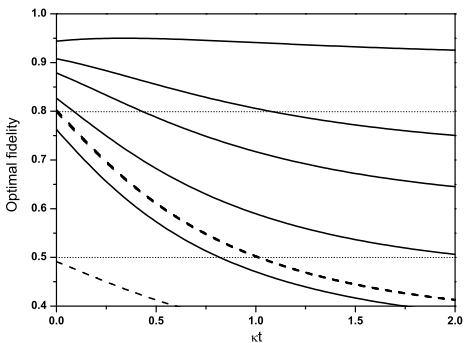}\caption{(Colour
online) The optimal fidelity for teleporting CSs as a function of $\kappa t$
with $\bar{n}=0$, $r=0.8,$ $T=0.9$ and $\chi=0.1,0.5,1,3,300$ from top to
bottom, respectively. For comparison, the fidelities with g=1 are plotted here
(dash lines).}%
\end{figure}

\section{Conclusions}

In this paper, we have examined the performance of three-tunable parameters in
realistic scheme of CV quantum teleportation with input coherent states and
the TMSV entangled resources. For our purpose, we have appealed to the input
and output relation in the CF formalism, which is convinent for nonideal
inputs and any entangled resources. In this realistic scheme, we have derived
the condition that the fidelity is independent of the amplitude of input
coherent states for any entangled resource. In order to investigate the effect
of three-tunable parameters on the fidelity of teleportation in the nonideal
scheme, we have derived the analytical expressions of the optimal fidelity for
input coherent states with three different prabability distributions and
investigated the performance of optimal fidelity. It is theoretically shown
that the usefulness of tunable parameters for improving the fidelity of
teleportation with or without the effect of environment and imperfect
measurements. In particular, for the input coherent states with a linear
distribution, the optimization with three tunable parameters is the best one
with respect to single- and two-parameter optimization, especially in the
region of small squeezing.

It would be interesting to extend the present analysis to teleport two-mode
states (ideal or nonideal cases) using multipartite (non-)Gaussian entangled
resources in the formalism of CF. In addition, a recent comparison between the
well-known CV VBK scheme and the recently proposed hybrid one by AR has been
made \cite{31,32}. It is shown that the VBK teleportation is actually inferior
to the AR teleportation within a certain range, even when considering a gain
tuning and an optimized non-Gaussian resource. Thus it may be worthy of
considering whether these three-parameter optimization can further improve the
fidelity in VBK scheme over that in AR scheme especially in the non-ideal scheme.

\textbf{ACKNOWLEDGEMENTS:} This work is supported by a grant from the Qatar
National Research Fund (QNRF) under the NPRP project 7-210-1-032. L. Y. Hu is
supported by the China Scholarship Council (CSC) and the National Natural
Science Foundation of China (No.11264018), as well as the Natural Science
Foundation of Jiangxi Province (No. 20151BAB212006).

\bigskip

\bigskip
\end{document}